# Multilayer network approaches to 'omics data integration in Digital Twins for cancer research

Hugo Chenel[1,2*], Malvina Marku[1], Tim James[3], Andrei Zinovyev[2*] and Vera Pancaldi[1*]

[1] CRCT, Université de Toulouse, Inserm, CNRS, Université Toulouse III-Paul Sabatier, Centre de Recherches en Cancérologie de Toulouse, Toulouse, France

[2] Evotec, Toulouse, France

[3] Evotec, Abingdon, Oxfordshire, UK

* Corresponding author: hugo.chenel@inserm.fr,
andrey.zinovyev@evotec.com,
vera.pancaldi@inserm.fr

## Abstract

This review examines current and potential applications of DTs in healthcare, focusing on the integration of multi-omics data using multilayer network approaches in cancer research. We discuss methodologies, tools and platforms commonly used for this integration, while highlighting case studies, challenges, and research gaps. Finally, we advocate the need for incorporating diverse data types to produce more effective DTs for personalization of cancer treatments and *in silico* clinical trials.

## Background

The emergence of precision medicine has shown the necessity to have integrative approaches that can exploit the vast and complex datasets generated by multi-omics technologies. These technologies, including (epi)genomics, transcriptomics, proteomics, metabolomics, metagenomics and other techniques that exhaustively characterize specific sets of biological entities, provides a comprehensive view of biological systems at multiple levels of regulation. Integrating these diverse types of datasets is essential to have a comprehensive digital representation of patients and identifying novel biomarkers or therapeutic targets (Hasin et al., 2017), especially in cancer, which will be the focus of this review.

Digital Twins (DTs) have emerged as a transformative paradigm in healthcare, providing dynamic, *in silico* models of cells, organs, or entire organisms that are continuously updated with real-time data. These virtual counterparts enable predictive modeling and personalized interventions (Kamel Boulos & Zhang, 2021). DTs leverage advanced computational techniques to integrate and analyze large-scale multi-omics data to implement precision oncology. A key feature of DTs is their use of network-based representations that span multiple scales and layers of biological information, often realized through multilayer networks. Multilayer networks ([Kivela et al., 2014](#)) have gained popularity for their ability to model complex systems where entities interact across multiple layers or dimensions. In the context of multi-omics data integration, these multilayer networks can represent different types of biological information as interconnected layers to capture the relationships and interactions between genes, proteins, metabolites and other molecular entities ([De Domenico, 2023; Trimbour et al., 2024](#)), enabling the identification of crucial regulatory mechanisms and pathways that may remain elusive in single-layer network analyses ([De Domenico, Lancichinetti, et al., 2015](#)). Multilayer networks can be thought of as a mathematical foundation of knowledge graphs ([Baptista et al., 2024](#)).

Despite the potential of DTs and multilayer networks, several conceptual and practical challenges in their implementation persist, including the intrinsic heterogeneous nature of multi-omics datasets, requiring elaborated algorithms and substantial computational resources. Ensuring data quality or consistency remains a significant obstacle. Moreover, the development of reliable DTs requires robust validation methods to ensure that their predictive capabilities align with real-world outcomes ([Hernandez-Boussard et al., 2021](#)).

## Digital Twins in precision health

### Definition and concept

By offering opportunities for individualized healthcare through virtual models, DTs represent a paradigm shift in the field of precision health and personalized medicine. In healthcare, DTs can be defined as highly detailed and dynamic virtual replicas of physical entities (cells, tissues, organs, patients and health systems), reproducing the structure, behavior and context of the physical counterpart ([Qi et al., 2021](#)). These virtual counterparts are continuously updated with real-time data to simulate and predict health outcomes, thereby optimizing clinical decision-making ([Fuller et al., 2020](#)). DTs integrate data from multiple sources in real-time to simulate health outcomes and can be adapted to each patient, potentially increasing proactive healthcare management by enabling more precise, timely and effective clinical interventions, ultimately improving patient treatment efficiency.

Beyond computational models, DTs are not limited to simple static digital models; they involve a bidirectional flow of information between the physical entity and its digital counterpart, ensuring that the virtual model evolves at the same time as its real-world counterpart. This continuous synchronization enables DTs to reflect the actual current state of the physical entity accurately, as well as predict future states under various scenarios ([Kamel Boulos & Zhang, 2021](#)). For instance, in clinical setting, DTs of a patient would integrate data from various sources (clinical records, genetic data, lifestyle information, real-time monitoring from wearable devices, …) to provide a comprehensive dynamic representation of the patient's health status and guide recommendations for therapeutic interventions or increased surveillance (**Figure 1**).

The effectiveness of DTs depends on their predictive capability ([Fuller et al., 2020](#)). Leveraging real-time data and historical health information helps DTs forecast various factors such as disease progression, outcomes of different treatment options or early signs of health deterioration. This predictive capability is particularly valuable in the management of chronic diseases, where early intervention can significantly influence disease progression. DTs can be used not only for improving disease treatments, but also for preventing their onset and mitigating their impact. Their predictive power comes from modeling disease trajectories, i.e. mapping a disease's progression over time, including its onset, development and chronicity, when relevant, starting from clinical records, multi-omics data and real-time health metrics. Therefore, they offer the potential for medical professionals to proactively adjust treatment plans, and anticipate significant turning points in the disease's course. They can also be used to provide alerts for early indicators of health deterioration for chronic diseases, whose trajectory frequently includes periods of remission or relapse.

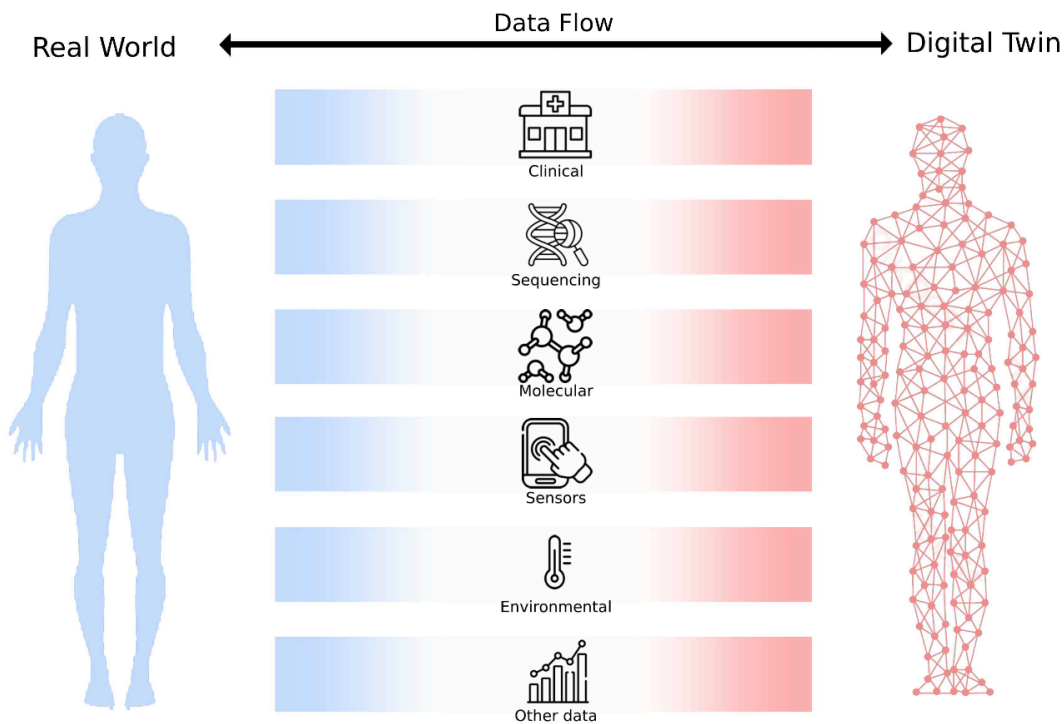

**Figure 1:** Conceptual diagram of DTs in the context of cancer research.

Moreover, the DTs framework can help in the creation of a learning healthcare system where continuous feedback from DTs informs clinical decisions and enhances healthcare delivery. This iterative process not only improves individual patient care, but also contributes to the broader knowledge base, driving key advancements in medical research and clinical practice (Kamel Boulos & Zhang, 2021).

## Potential applications in healthcare

Personalized medicine and disease management are a few of the areas in which DTs are gaining recognition for their transformative potential in the healthcare industry. Several existing systems integrate multi-omics data with clinical and lifestyle information to create a more complete and individualized health profile that can serve as a basis for *in silico* disease modeling. These models can form the basis for clinical decision support systems, providing predictions of responses to the different treatment options, a highly valuable tool for clinicians who can tailor interventions based on the unique characteristics of each patient. In addition, DTs enable customized interventions by utilizing real-time data to offer precise and dynamic insights into patient health but they involve continuous mapping of the physical and digital counterparts.

Virtual representations play an important role in the prevention and management of chronic disease. The constant monitoring of physiological data allows DTs to identify early signs of disease onset but also aggravation, which is crucial for timely intervention. In the context of diabetes management, DTs have demonstrated the capacity to predict glucose fluctuations by integrating data on dietary intake, physical activity or medication adherence, making it a powerful tool to help clinicians with better management of the disease and preventing complications (Shamanna et al., 2021). Recently, this has led to the design of DT systems that inject insulin through real-time monitoring of patient glucose levels (Thamotharan et al., 2023). Other studies in cardiovascular health demonstrate how DTs can monitor vital or relevant cardiovascular signs (resting heart rate, heart rate variability, blood pressure) to detect abnormalities at an early stage and suggest preventive measures (Coorey et al., 2022).

Specifically in oncology, advanced computational frameworks have proven useful to model tumor growth and predict how a tumor will respond to different therapeutic interventions (Benzekry et al., 2014) leading to optimization of cancer patient treatment plans by improving drug effectiveness and minimizing side effects (Bruynseels et al., 2018).

Only a few real DTs for oncology have been developed, also due to the difficulty of dynamically adapting treatments. Some studies demonstrated a significant increase in treatment efficacy that shows the potential of DTs in adapting cancer therapies, each simulation predicting a trajectory for the patient's cancer under one of the treatment options (Hernandez-Boussard et al., 2021). Additional examples of oncological DTs are reviewed in (Stahlberg et al., 2022), illustrating the limits but also the potential of DTs to improve early diagnosis of cancers that are generally detected at advanced stages. One pilot project generated virtual patients for early detection of pancreatic cancer. Subtle changes in biomarker profiles could be detected at an early stage, which is indicative of early pancreatic cancer. However, matching real patients in the cohort with simulated patients from the model did not result in a definitive mapping between real-world data and model parameters. Another project involved using simulations of cellular interactions in the tumor to personalize melanoma treatment.

Another remarkable application predicts drug efficacy in lung cancer using DTs. DTs models allow the simulation of different delivery methods and dosages under different therapeutic scenarios to establish treatment efficacy. Understanding these mechanisms is essential for clinicians to be able to adjust treatment plans as a therapeutic or preventive measure (Vallée, 2024). Some studies also underscored the DTs value in the anticipation of treatment response in cancer therapy (Venkatesh et al., 2024).

On the other side, further from real patients, testing new drugs or treatments via virtual platforms is an application of DTs that is transforming drug discovery. The simulation of biological processes and disease progression is implemented in DTs to predict the interaction of new compounds with different molecular targets, observing how it will perform in simulated diverse patient populations constituting virtual cohorts (H. Wang et al., 2024). This approach reduces the need for extensive *in vivo* testing, improving the drug development process. DTs could also play a major role in patient education by improving their own awareness, and pushing them to be proactive about their health. If an interactive representation of health status via Patient Reported Outcomes and the impact of different lifestyle choices is provided to patients, DTs could indirectly enhance motivation for self-management. For instance, DTs can show some patients with obesity the impact of weight loss on some health metrics, encouraging changes in diet and exercise regimens (Gkouskou et al., 2020)*.*

In conclusion, the applications of DTs in healthcare are broad: personalized medicine, disease prevention/management, drug discovery and patient engagement. The integration of real-time data by DTs offers dynamic insights that ameliorate personalized healthcare. As the technology continues to evolve, the potential of these DTs in improving healthcare outcomes should continue to grow, highlighting their importance in the future of precision health (Fuller et al., 2020; Kamel Boulos & Zhang, 2021). Literature examples reveal the potential of dynamic and predictive models to improve early detection, personalize treatment strategies or therapeutic resistance anticipation. However, the development of tools that enable feedback between the DT and real patients is still in progress (Fuller et al., 2020), holding promise in cancer research.

## Challenges and opportunities in the integration of multi-omics data in cancer research

A full comprehension of molecular mechanisms driving cancer progression is made possible by the integration of multi-omics data, which has become an essential component in oncology. The word "omics" refers to a group of scientific areas, each of which focuses on specific biological entities (Srivastava et al., 2024). Genomics, epigenomics, transcriptomics, proteomics, metabolomics, metagenomics and even radiomics (medical imaging such as MRI, CT scans, …) are the main categories of data that are important for cancer research. We give an overview of different types of 'omics data, their general characteristics and relevance to cancer research in Table 1.

The integration of multi-omics data in oncology offers significant potential to advance the understanding of cancer biology. However, several significant challenges must be addressed to fully realize its potential. These challenges include technical, computational and biological nature, as well as data administration policies, while also presenting unique opportunities for innovation.

One of the main challenges in multi-omics integration is the inherent heterogeneity of the data, stemming from both its nature and its method of generation. Different 'omics data sources are generated using diverse technologies, each having its own data formats (Cao et al., 2021; Ren et al., 2023), indicating the necessity for a deeper investigation of the crosstalk between different levels of data granularity (Cirillo et al., 2021). This inconsistency adds complexity to the integration process because the harmonization and standardization of these datasets require advanced methodologies (Hasin et al., 2017).

**Table 1:** Different types of 'omics data, their characteristics and relevance to cancer research.

| 'Omics Type | Description | Technologies | Applications to cancer research |
|---|---|---|---|
| **Genomics** | Study of the complete DNA sequence, identifying mutations and genetic alterations driving disease | Whole-genome sequencing (WGS), Whole-exome sequencing (WES) | Identification of oncogenes, tumor suppressor genes, and genetic drivers of cancer. |
| **Epigenomics** | Study of changes in gene expression that do not involve changes in the DNA sequence, including DNA methylation and histone modifications. | ChIP-seq, DNA methylation assays | Understanding gene silencing, tumor suppressor inactivation, and biomarker identification for cancer detection and prognosis. |
| **Transcriptomics** | Study of the complete set of RNA transcripts, providing insights into gene expression and regulatory mechanisms. | RNA sequencing (RNA-seq) | Identifies dysregulated genes, tumor heterogeneity, and potential biomarkers. |
| **Proteomics** | Large-scale study of proteins, including their expression, modifications, and interactions. | Mass spectrometry (MS), Protein microarrays | Identification of altered protein expression, protein biomarkers, and targets for therapy in cancer. |
| **Metabolomics** | Study of metabolites, small molecules that are intermediate products of metabolism, offering a functional readout of biochemical activities. | Nuclear magnetic resonance (NMR), Mass spectrometry (MS) | Understanding metabolic alterations in cancer, such as the Warburg effect, and linking metabolism to tumor development. |
| **Metagenomics** | Analysis of the genetic material from microbial communities, offering insights into the microbiome's role in cancer. | Shotgun metagenomic sequencing, 16S rRNA sequencing | Exploration of microbiome influence on tumorigenesis and potential microbiota-based cancer therapies. |
| **Spatial 'Omics** | Study of the spatial organization of cellular components within the TME, preserving tissue architecture during analysis. | Spatial transcriptomics, spatial proteomics | Visualization of cellular heterogeneity and spatial patterns in tumor ecosystems. |
| **Radiomics** | Extraction of quantitative features from medical imaging data (e.g., MRI, CT, PET), analyzing imaging patterns to predict tumor phenotype and treatment response. | MRI, CT, PET, Advanced image processing algorithms | Enables the non-invasive prediction of tumor characteristics, prognosis, and response to therapies by analyzing imaging data. |
| **Exposomics** | Study of lifetime environmental exposures and their interaction with the genome, focusing on | Data mostly from questionnaires or sensors | Provides insights into how environmental factors contribute to cancer risk |

| | external and internal factors such as chemicals, diet, and lifestyle. | | through gene-environment interactions. |
|---|---|---|---|

The high volume of multi-omics data requires efficient algorithms for extracting useful and interpretable information. Also, this high-dimensional data integration involves handling large-scale datasets that demand high computational power as well as scalable storage capacity (Nikolski et al., 2024). Moreover, the development of efficient algorithms for the different steps of data processing (preprocessing, normalization and integration) is crucial to managing this complexity (Santiago-Rodriguez & Hollister, 2021). Machine learning (ML) or artificial intelligence (AI) techniques are increasingly employed to address these challenges but their implementation needs substantial expertise and resources (Angermueller et al., 2016; Libbrecht & Noble, 2015). Ensuring data quality and reproducibility is another significant challenge. 'Omics data can be affected by various technical issues at different steps of the experiments or analysis pipeline (technical biases, batch effects, sample-specific variations, …) which can hide biological signals and lead to false conclusions (Leek et al., 2010). Therefore, rigorous quality control measures, standardized protocols and reproducible workflows are essential to mitigate all these issues. Additionally, the integration process must account for missing data and varying sample sizes across different 'omics layers (Kim et al., 2005).

Complex datasets are generated by the integration of multi-omics data, which requires careful interpretation to derive biologically meaningful insights. For instance, the identification of key drivers in cancer and the elucidation of their functional roles in the context of multi-omics data is an important and challenging task. Distinguishing true biological signals from noise and correlating multi-omics findings with clinical outcomes are the true challenges. Collaborative efforts between researchers and clinicians are essential to contextualize the results and then translate them into actionable knowledge (Demirel et al., 2022).

Despite all these challenges, the integration of multi-omics data currently offers substantial opportunities to deepen our understanding of cancer biology in the tumor site and across the whole organism. The combination of information from several types of 'omics data can bring an extensive view of the molecular mechanisms that drive tumor growth via different types of data integration (S. Huang et al., 2017) (**Figure 2**). This global approach helps to identify new biomarkers that may not be apparent from single-omics studies alone, with some methods taking into consideration data type diversity and big data volume (D. Wu et al., 2015).

Significant potential in personalized medicine and great progress can be achieved thanks to multi-omics integration. The detailed molecular profiling of individual tumors is adequate to reveal patient-specific vulnerabilities, thus designing customized therapeutic strategies for the patients (Collins & Varmus, 2015). For instance, integrating genomic and transcriptomic data is interesting to identify actionable mutations or dysregulated pathways, while proteomic and metabolomic analyses can improve our understanding of tumor metabolism alterations (Gegner et al., 2023; W. Zhang & Plevritis, 2018).

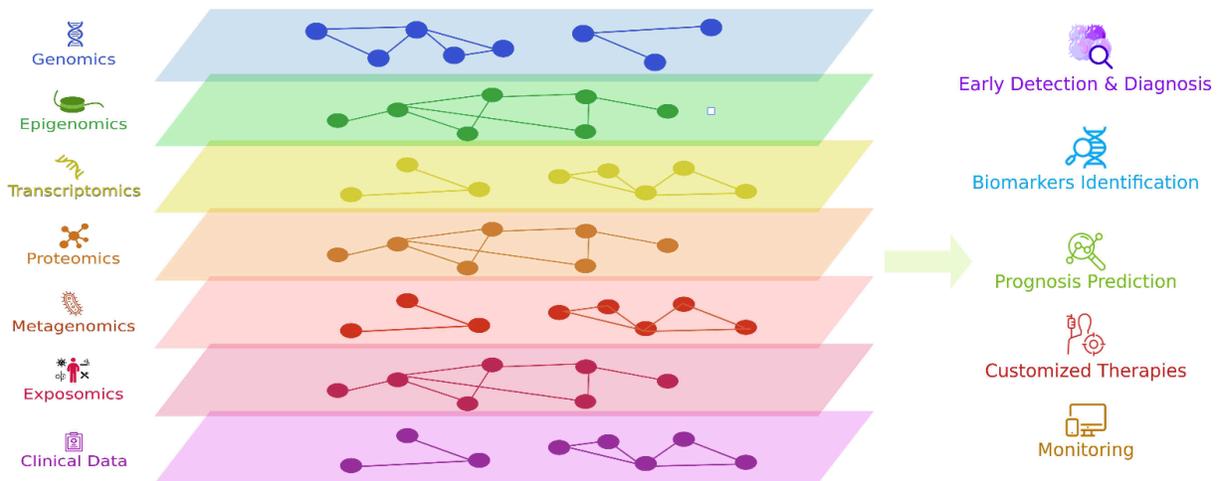

**Figure 2:** The dynamic interactions of multi-omics layers and their impact on therapeutic strategies (treatment or prevention).

Developing novel algorithms or ML models to handle high-dimensional, heterogeneous data is another emergent area of research (C. Wu et al., 2019). These advancements not only improve the performance (efficiency and accuracy) of data integration but also contribute in a broader way to the field of computational biology. The complexity of the integration makes collaborative efforts and data sharing among researchers and institutions absolutely necessary (Gómez-López et al., 2019). Some initiatives such as The Cancer Genome Atlas (TCGA) (Weinstein et al., 2013), International Cancer Genome Consortium (ICGC) (J. Zhang et al., 2011) and Human Cell Atlas (HCA) (Rozenblatt-Rosen et al., 2017) provide highly valuable resources and infrastructure for multi-omics research. Some recent resources are specific to certain types of 'omics data such as the Human Protein Atlas (Pontén et al., 2008) (proteomics), International Human Epigenome Consortium (IHEC) (Stunnenberg et al., 2016) (epigenomics) or Human Metabolome Database (HMDB) (Wishart et al., 2007) (metabolomics). These collaborative platforms are excellent for the exchange of data, tools or expertise, accelerating scientific discoveries and clinical translation.

## Multilayer network approaches for multi-omics integration

### Fundamentals of multilayer networks

Multilayer networks are a powerful framework to model complex systems characterized by multiple types of interactions (or relationships). Unlike traditional single-layer networks which represent a single type of interaction among molecular entities, multilayer networks incorporate multiple interconnected layers, where each of them represents a different type of relationship, capturing the multifaceted nature of real-world interactions (De Domenico, 2022; Kivela et al., 2014). More specifically, a multilayer network is formally defined by a set of nodes and layers, where each node can exist in one or more layers, while edges represent the connections between nodes within the same layer (intra-layer edges) or across different layers (inter-layer edges) (Figure 3). The interconnections between the different layers (inter-layer edges) are key elements in the network to understand the influence of different types of interactions and the impact on the overall dynamics of the system.

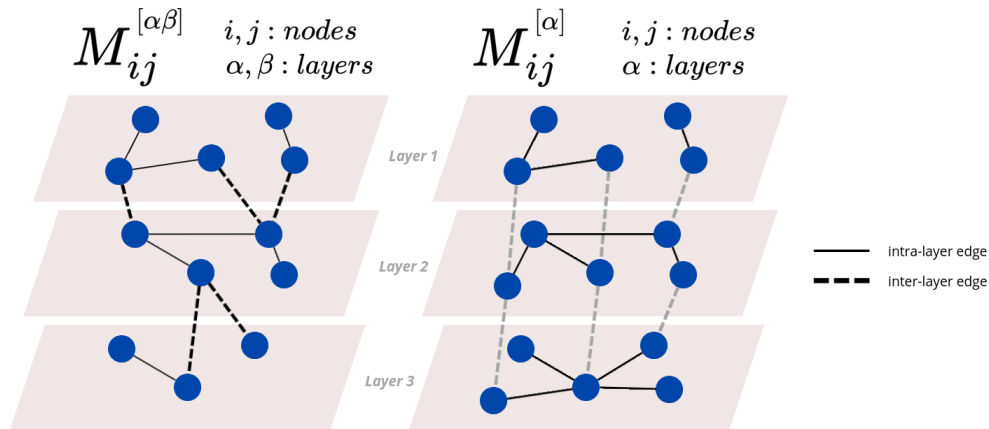

**Figure 3:** Schematic representation of multilayer and multiplex networks. On the left, a multilayer network showing intra- and inter-layer connections. On the right we represent a special case of multilayer, a multiplex network, in which the nodes are the same in all layers.

One of the main advantages of multilayer networks, compared to single-layer networks, comes from their ability to model interdependencies between different types of data. In system biology research, a multilayer network can simultaneously represent genomic, transcriptomic, proteomic and clinical data, each as a distinct layer. The interactions within each layer are able to capture relationships specific to that data type, such as transcription factors (TFs) - target genes (TGs) regulatory interactions in the transcriptomic layer or protein-protein interactions (PPI) in the proteomic layer. On the other hand, inter-layer edges capture the relationships between different data types (De Domenico, 2018). The construction of multilayer networks involves several methodological steps:

1. **Data preprocessing to ensure compatibility:** Ensures consistency across 'omics datasets, involving quality control, noise reduction, and standardization. Batch effect removal and data imputation address technical biases and missing values, to accurately reflect biological variability (Lazar et al., 2013; Leek et al., 2010).
2. **Single-layer network reconstruction:** For each 'omics type, individual single-layer networks are constructed, based on annotated interactions such as TF-target interactions, PPI, or metabolic pathways. Algorithms tailored to each data type can establish connections indicating correlation, causality, or interaction strength, with edge direction or weight based on interaction type and confidence.
3. **Integration into the multilayer network:** Single-layer networks are integrated into a multilayer network via inter-layer edges, representing cross-'omics interactions, i.e. regulatory relationships between genes and proteins, or other molecular entities. Several approaches can be used to define inter-layer edges, including data-driven methods that infer relationships based on statistical correlations or ML models that predict interactions based on prior knowledge of biological principles. A special case of multilayer networks is obtained when all nodes are present in all layers denoted as multiplex networks, where distinct layers represent different interaction types for the same set of nodes, preserving the individual network topologies (Kivela et al., 2014; Valdeolivas et al., 2019).

The visualization of the constructed network can be done using *ad hoc* developed tools, or extensions of other tools initially built for single-layer networks visualization, which we summarize in Table 2.

**Table 2:** Tools used to visualize multilayer networks.

| Tool | Description |
| --- | --- |
| **MUXViz** (De Domenico, Porter, et al., 2015) | Visualization of multiplex networks, allowing the exploration of multiple layers of interactions in the same system. |
| **Omics Integrator** (Tuncbag et al., 2016) | Incorporation of different 'omics data into coherent networks, optimized for gene, protein, and pathway interactions. |
| **NetZoo** (Ben Guebila et al., 2023) | Specialization in network-based ML for analyzing large, multi-omic datasets, with a focus on uncovering hidden structures in biological networks. |
| **Cytoscape** (Shannon et al., 2003) **NetworkX** (Hagberg et al., 2008) **iGraph** (Csárdi et al., 2006) **Gephi** (Bastian et al., 2009) | While initially built for single-layer networks, these tools offer flexible network visualization capabilities, which can be adapted for multilayer network exploration. |

**Analysis of multilayer networks**

The advancement of computational tools, algorithms and graph-based methods is essential to analyze multilayer networks and interpret their features. One approach is the use of centrality measures to determine the importance of nodes within the multilayer network. Frequently used centrality measures for single-layer networks (gene degree, betweenness centrality, alpha centrality, …) are extended to consider the multilayer context, offering the opportunity to find nodes that play a major role across multiple layers (De Domenico, Lancichinetti, et al., 2015). Another essential technique is community detection, an analytical technique which involves node cluster (or community) identification; these nodes are densely connected amongst each other within and across layers compared to the connections they have with other nodes not in the community. It helps to reveal functional modules that work together in a coordinated way in several contexts (Didier et al., 2015; Mucha et al., 2010), particularly in complex diseases (Choobdar et al., 2019; Halu et al., 2019). Additionally, examining multilayer motifs means looking at small recurring patterns of interconnections that occur within and between layers of the network. These motifs establish the fundamental building blocks of the network's structure (Boccaletti et al., 2014). More advanced concepts, such as the Von Neumann entropy of a multilayer network, can be utilized to characterize global connectivity patterns across layers (De Domenico, Nicosia, et al., 2015). Studying network dynamics is also interesting to understand how information propagates through the multilayer network (a phenomenon called flow) via modeling the spread of perturbations and understanding how interactions in one layer affect dynamics in other ones (De Domenico et al., 2016; Gómez et al., 2013). One of the most used for this purpose approach is Random Walk with Restart, which identifies relationships between nodes, reconstructs communities, and predicts key interactions (Baptista et al., 2024; Cowen et al., 2017).

**Machine learning and predictive modeling**

ML techniques are increasingly applied to multilayer networks to develop predictive models. Supervised learning algorithms, like support vector machines (SVM) or random forests (RF), can be trained on multilayer network features to predict clinical outcomes or response to therapy (Libbrecht & Noble, 2015). Unsupervised learning methods (clustering or dimensionality reduction) can help to discover new patterns or possibly some associations within the integrated data (C. Wu et al., 2019). Predictive modeling using multilayer networks enhances the ability to make data-driven decisions in personalized cancer treatment.

Network embedding approaches are also effective in analyzing a wide range of networks, even if they were not originally developed for biological networks (Parvizi et al., 2022). These methods have shown good performance in tasks like community detection, node classification and link prediction. However, most embedding methods are not designed to be applied on multiplex networks. To address this, advanced methods like MultiVERSE have been developed; it has demonstrated superior performance in tasks such as link prediction and network reconstruction, particularly in multiplex networks (Pio-Lopez et al., 2021).

With the ability to learn complex relationships and inherent structures in multilayer networks, Graph Neural Networks (GNNs) are highly effective at analyzing graph-structured data (Corso et al., 2024). They encode both the node features and the graph topology into low-dimensional embeddings; these embeddings can then be used for a variety of tasks, such as node classification and link prediction (M. Zhang, 2022). Latest advancements, like Graph Convolutional Networks (GCNs) and Graph Attention Networks (GATs) enhance the capacity to represent heterogeneous data and facilitate uses such as link prediction applied to drug target prediction (Chatzianastasis et al., 2023; Chen et al., 2024; Grassia et al., 2021; K. Huang et al., 2024).

**Applications of multilayer networks in data integration**

Data integration based on multilayer networks is performed by the inference or construction of integrated networks that combine multi-omics (genomic, transcriptomic, proteomic, metabolomic, etc.) data. This all-around approach is ideal to map the interactions between different molecular entities across various biological layers, leading to a multidimensional view of cellular processes. We can consider that genomic data can reveal mutations in cancer-related genes, while transcriptomic data can show how these mutations affect gene expression levels. Integrating these datasets into a multilayer network (Trimbour et al., 2024) could be a way to establish the influence of genetic alterations on transcriptional changes and possibly the propagation of these changes through proteomic and metabolomic layers.

A few diverse multilayer networks have been constructed, for instance integrating disease-perturbed proteins, drug targets and biological functions into a multi-level interactome network for a better definition of disease mechanisms (Ruiz et al., 2021), while similar approaches employing multilayers have also been applied to the study of specific pathologies, such as infection by Sars-Cov-2 (Verstraete et al., 2020).

In cancer research, significant benefits emerge from the application of multilayer network theory, with the central one coming from the integration of diverse types of biological data that enhance the view of the molecular mechanisms driving tumors. This integrative approach can help to identify key regulatory pathways and potential therapeutic targets that might be overlooked in single-layer analyses. Overall, multilayer networks can help to interpret the complex interplay between genetics and environment (genetic mutations, epigenetic modifications, environmental factors, …) in cancer evolution.

## Digital Twins for cancer research: a multilayer network perspective

### Conceptual framework

The conceptual framework for utilizing DTs in oncology through a multilayer network perspective covers a systematic approach that integrates diverse types of biological data to create predictive models of tumors. This framework leverages the positive aspects of multilayer network theory to deliver a more complete representation of the biological processes and molecular functions involved in cancer progression (Laubenbacher et al., 2022). In the following sections, we consider the multilayer network as the scaffold of a patient DT and give a roadmap of how to build it by integrating multi-omics data, as well as how to use it in cancer research.

### Integration of multi-omics data

The integration of multi-omics data is the central area of this conceptual framework, which can include genomics, epigenomics, transcriptomics, proteomics, metabolomics and eventually other data types like clinical data (e.g. histopathology images), generated from tumor samples, as well as data regarding the host (metagenomics of the microbiome, exposomics etc). Each of these layers brings unique and complementary insights into the different aspects of cancer.

### Clinical implementation and validation

The final step in the framework corresponds to the clinical implementation and validation of the DT, a challenging but primordial task for the technology to gain credibility. In this context, two different concepts exist. The first one corresponds to the integration of DTs into clinical workflows where they can assist oncologists in the decision-making process with data-driven choices regarding diagnosis, prognosis or treatment (clinical decision support systems). The second one is based on the use of DTs to perform virtual clinical trials, testing the efficacy of new compounds or even drug associations in a virtual environment before applying them to patients to limit the risk (H. Wang et al., 2024).

Validation of DTs' predictive capabilities is indispensable to ascertain their reliability. Some retrospective analysis using historical patient data, as well as prospective validation in clinical trials can be performed. Continuous monitoring and updating of DTs with dynamic, more recent patient data reinforce the accuracy and confirm the effectiveness of the tool for cancer management (Hernandez-Boussard et al., 2021).

## Practical Challenges in DT implementation

### Research gaps

The availability of longitudinal data to capture the dynamic nature of tumor progression is a significant challenge. This causes a limited understanding of temporal dynamics in cancer progression or even treatment response. The continuous monitoring and data acquisition in real-time from patients involves the integration of clinical, imaging or wearable device data, as well as, potentially, patient-reported outcomes, with multi-omics profiles (Chow et al., 2024). Achieving this level of integration requires advanced data management systems that are capable of handling diverse data flows but also maintaining temporal consistency.

A very important issue that should not be neglected is the use of sensitive personal health information (privacy, consent, data ownership) because it raises some privacy and security concerns. Among them, protecting patient data from unauthorized access is critical as data breaches can lead to potentially severe consequences like identity theft or loss of trust in healthcare systems (Rothstein, 2015; Thapa & Camtepe, 2021). Data privacy should be made a priority by implementing stringent data governance policies to protect patient information. Techniques such as differential privacy, which adds noise to datasets to prevent the identification of individuals are more and more employed to protect patient confidentiality while allowing data analysis (Dwork & Roth, 2014). More recent solutions are promising like blockchain, which offers a decentralized and immutable ledger that can be used to securely store/share patient data (Pournaghi et al., 2020), or homomorphic encryption, allowing computations on encrypted data without needing to decrypt it first (Munjal & Bhatia, 2023). Security measures must also address the risk of cyber-attacks with the implementation of robust cybersecurity protocols (regular security audits, intrusion detection systems, secure access controls, …) which is essential to protect multi-omics data and DT models from hostile attacks (Fernández-Alemán et al., 2013).

The biological interpretation of integrated multi-omics data is another research gap. While multilayer networks can reveal complex interactions, translating these findings into concrete biological insights remains a challenge (Demirel et al., 2022). There is a need for other tools to interpret the results of multilayer network analyses in not only a technically and biologically meaningful way. In addition, experimental validation of the predictions made by DTs is necessary to confirm different parameters like accuracy or utility in clinical settings.

### Data collection and AI

The adoption of DTs in precision medicine is increasingly operational, driven by extensive data collection alongside traditional biomedical methodologies. However, the reliance on black-box predictive models based on large datasets presents limitations that could potentially slow down the broader application of DTs in clinical settings. In the literature, it has been argued that hypothesis-driven generative models (generate data based on a set of underlying assumptions of the processes that generate the observed data), more particularly multi-scale modeling, are essential to boost the clinical accuracy of DTs. The transformative potential of DTs in healthcare has been explored by emphasizing their capability to simulate complex interdependent biological processes across multiple scales. The integration of generative models with extensive datasets can deliver scenario-based modeling approaches to explore diverse therapeutic strategies. It is an excellent strategy to support dynamic clinical decision-making. This method not only leverages advancements in data science to improve disease management but also incorporates insights from complex systems, quantitative biology and digital medicine to improve patient care (De Domenico et al., 2024).

## Perspectives

In the context of cancer, multilayer networks have been used to capture complex interactions, leading to the identification of novel therapeutic targets (Liu et al., 2020). The integration of gene expression data with chromatin accessibility and protein-protein interaction networks was shown to uncover key regulatory proteins that are central to tumor progression, which might not be clearly identified when examining single-layer networks, highlighting the added value of a multilayer approach to propose new biomarkers.

Given the importance of the TME in affecting tumor progression and therapy response (Quail & Joyce, 2013), it is imperative to explore the integration of information regarding all cellular populations within a sample into predictive models. This process requires the non-trivial development of new methods that capture salient information from either estimation of cell type proportions from bulk RNAseq (Merotto et al., 2024) or from single-cell spatial and non-spatial datasets. It is likely that new modes of dimensionality reduction will have to be applied to these types of data to produce correspondences between these omics and more traditional layers, in which nodes are molecular entities.

In fact, multi-level and hybrid models will probably become necessary to use the multilayer structures beyond simple data integration and towards dynamic modeling. Despite the obvious interest of analyzing the integrated datasets on this multilayer structure, looking for communities spanning several layers, for example, the real power of this framework will come from the construction of dynamic models of these complex systems. A first approximation of dynamics can be provided by flow analyses on the network, which simulate how the information will spread within and across layers, providing an idea of perturbations that can move the DT into a required state (for instance towards a healthy state). Simulating the dynamic state of a patient, including the tumor itself, the immune system, different organs and how the patient environment is impacting all of them will probably require a combination of models of different scales (e.g. molecular, metabolic, cellular, etc.) as well as a better understanding of regulation and correspondence between nodes in different layers. Further down the line, with a better understanding of the underlying multi-level biological processes, we might be able to construct executable models comprising the entire variety of datasets, constituting DTs that will be extremely useful for in-silico clinical trials that consider the patient as a whole.

Several approaches to modeling the TME have been proposed (Johnson et al., 2023) and these might enable a dynamic perspective, essential to understanding the adaptation of cancer cells to therapeutic molecules at the tumor site. More effective treatment strategies might require combinations of different approaches at different times, requiring a good understanding of the time scales of the different biological processes involved in cancer (Z. Wang & Deisboeck, 2019). To move closer to virtual clinical trials, a more complete representation of patients would be required. The flexibility of the multilayer approach should enable this by allowing the integration of even more diverse data types. For instance, integrating prior comorbidities could be of great value in predicting patient outcomes, as was the case for predicting pancreatic cancer onset (Placido et al., 2023). Indeed patient comorbidity profiles might work as proxies for identifying patient characteristics that are relevant for patient stratification (Sánchez-Valle et al., 2020), while integrating exposures or side effects with tumor characteristics via multi-layer networks and multi-level models could produce more useful DTs for drug discovery and development in *in silico* clinical trials.

## Conclusion

This review has examined the methodologies, tools and challenges in the implementation of DTs for cancer by their potential to provide a comprehensive view of tumor biology, a central element in cancer research for precise predictions of disease progression, response to treatment and side effects. Despite the hopes in DTs, some challenges (data integration, computational efficiency or data privacy) need to be faced. Future research should focus on developing computational models and enhancing scalability without neglecting ethical frameworks for patient data use. Collaboration between industrialists, clinicians and researchers could help overcome these challenges. DTs offer opportunities to transform cancer care with more effective and individualized treatments by providing broad dynamic models of cancer. Further advancements in this field, with the integration of more patient-level data, are crucial, holding the promise of enabling better *in silico* clinical trials and accelerating cancer treatment development.

## Author contribution

Conceptualization: H.C; Writing: H.C, M.M, A.Z, V.P; Reviewing: H.C, M.M, T.J, A.Z, V.P; Supervision: V.P, A.Z.

## Competing interests

The authors have declared that no competing interests exist.